\documentclass{jetpl}
\twocolumn
\usepackage{graphicx}
\lat

\title{Sensitivity of Baikal-GVD neutrino telescope to neutrino emission toward the center of Galactic dark matter halo}

\rtitle{Sensitivity of Baikal-GVD neutrino telescope to neutrino emission toward the center of Galactic dark matter halo\ldots}

\sodtitle{}

\author{A.D.\,Avrorin$^{a}$, A.V.\,Avrorin$^{a}$,
  V.M.\,Aynutdinov$^{a}$, R.\,Bannasch$^{g}$,
  I.A.\,Belolaptikov$^{b}$, D.Yu.\,Bogorodsky$^{b}$,
  V.B.\,Brudanin$^{b}$, N.M.\,Budnev$^{c}$, I.A.\,Danilchenko$^{a}$,
  S.V.\,Demidov$^{a}$\/\thanks{demidov@ms2.inr.ac.ru},
  G.V.\,Domogatsky$^{a}$, A.A.\,Doroshenko$^{a}$, A.N.\,Dyachok$^{c}$,
  Zh.-A.M.\,Dzhilkibaev$^{a}$, S.V.\,Fialkovsky$^{e}$,
  A.R.\,Gafarov$^{c}$, O.N.\,Gaponenko$^{a}$, K.V.\,Golubkov$^{a}$,
  T.I.\,Gress$^{c}$, Z.\,Honz$^{b}$, K.G.\,Kebkal$^{g}$,
  O.G.\,Kebkal$^{g}$, K.V.\,Konischev$^{b}$, E.N.\,Konstantinov$^{c}$,
  A.V.\,Korobchenko$^{c}$, A.P.\,Koshechkin$^{a}$,
  F.K.\,Koshel$^{a}$, A.V.\,Kozhin$^{d}$, V.F.\,Kulepov$^{e}$,
  D.A.\,Kuleshov$^{a}$, V.I.\,Ljashuk$^{a}$, M.B.\,Milenin$^{e}$,
  R.A.\,Mirgazov$^{c}$, E.R.\,Osipova$^{d}$, A.I.\,Panfilov$^{a}$,
  L.V.\,Pan'kov$^{c}$, A.A.\,Perevalov$^{c}$, E.N.\,Pliskovsky$^{b}$,
  M.I.\,Rozanov$^{f}$, V.F.\,Rubtzov$^{c}$, E.V.\,Rjabov$^{c}$,
  B.A.\,Shaybonov$^{b}$, A.A.\,Sheifler$^{a}$, A.V.\.Skurihin$^{d}$,
  A.A.\,Smagina$^{b}$,
  O.V.\,Suvorova$^{a}$\/\thanks{suvorova@cpc.inr.ac.ru},
  B.A.\,Tarashansky$^{c}$, S.A.\,Yakovlev$^{g}$,
  A.V.\,Zagorodnikov$^{c}$, V.A.\,Zhukov$^{a}$, V.L.\,Zurbanov$^{c}$\\~\\}

\rauthor{A.D.\,Avrorin et al.}

\sodauthor{A.D.\,Avrorin et al.}

\address{$^{a}$Institute for Nuclear Research RAS, 
117312 Moscow, Russia\\~\\
$^{b}$Joint Institute for Nuclear Research, Dubna, Russia\\~\\
$^{c}$Irkutsk State University, Irkutsk, Russia\\~\\
$^{d}$Skobeltsyn Institute of Nuclear Physics MSU, Moscow, Russia\\~\\
$^{e}$Nizhni Novgorod State Technical University, Nizhni Novgorod, Russia\\~\\
$^{f}$St.Petersburg State Marine University, St.Petersburg, Russia\\~\\
$^{g}$EvoLogics GmbH, Berlin, Germany}

\abstract{We analyse sensitivity of the gigaton volume telescope
  Baikal-GVD for detection of neutrino signal from dark
  matter annihilations or decays in the Galactic Center. Expected
  bounds on dark matter annihilation cross section and its
  lifetime are found for several annihilation/decay channels.  
}

\PACS{74.50.+r, 74.80.Fp}

\begin{document}

\maketitle

There is a lot of evidence on existence of dark matter
  (DM). Cosmological observations indicate that DM contributes about
  27\% in total energy density within $\Lambda$CDM model. Weakly
  Interacting Massive Particles (WIMP) are among the most interesting
  candidates for the dark matter. Currently, tremendous efforts are
  put in the searches for DM at colliders, in direct detection
  experiments and, finally, in indirect searches for a signal in
  products of WIMP annihilations/decays in astrophysical
  observations. The Galactic Center (GC) is a very promising region to
  look for such a signal. Recent analysis of gamma telescope FERMI-LAT  
  dataset performed by few groups for several years of observation
  indicates on the existence of a diffuse gamma-ray excess from the
  center of our Galaxy at energies 10--20 GeV and a gamma-line
  feature about 133 GeV with a significance about
  3$\sigma$~\cite{FERMI-Mazziotta:2014}. 
  There are many systematic uncertainties here due to gamma-ray
  background from galactic diffuse emission and from close around
  local astrophysical sources like pulsars or supernova remnants. 
Another attractive possibility is to look for neutrino signal from DM
and it has been put forward by neutrino telescope collaborations. 
Recently the IceCube collaboration announced four dozen
candidates~\cite{ICubeAstro:2013} for neutrinos of astrophysical
origin with energies above hundreds TeV. However there is no any
significant clustering of events in any direction. 

New Baikal-GVD project aimed on installation of gigaton volume
detector in lake Baikal~\cite{Baikal:2009,Baikal:2011,GVD:2013} is now
in progress. The main goal of GVD telescope is a search for neutrinos
of astrophysical origin. Here, we estimate one year sensitivity to
look for  possible neutrino signal from dark matter in the center of our
galaxy for the GVD telescope in planned configuration of 12
clusters composed by 2304 photodetectors per 96 strings. The telescope
GVD is building in a place of previous telescope
NT200~\cite{NT200:1997} in the south basin of the Lake Baikal, at a 
distance 3.5 km off the shore. Local coordinates are 51.83$^\circ$ N
and 104.33$^\circ$ E. Position provides average visibility
for the GC almost 75\% per day. Actual position of the center of
Galaxy to be used is taken with right ascension $\approx
266.31^\circ$ and declination $\approx-29.49^\circ$. 

The Baikal water optical properties have been studied for a long time
and are characterized by absorption length $20\div24$ m at 480 nm and
and a scattering length $30\div70$ m depending on season. The total
trigger rate is expected to be approximately 100 Hz, dominated by
downgoing atmospheric muons. The detection of relativistic particles 
crossing effective volume of a deep underwater telescope implies  
collection of their Cherenkov radiation by optical modules (OMs)
synchronized in time and calibrated in pulses. Events arriving
from down hemisphere are considered as candidates on those originated
from  neutrino scatterings off nucleons in surrounding water or
rock. The main challenge is to suppress atmospheric muon background
from upper hemisphere exceeding upward going neutrino flux by factor $10^6$. 

\begin{figure}
\begin{center}
\includegraphics[width=0.33\textwidth,angle=0]{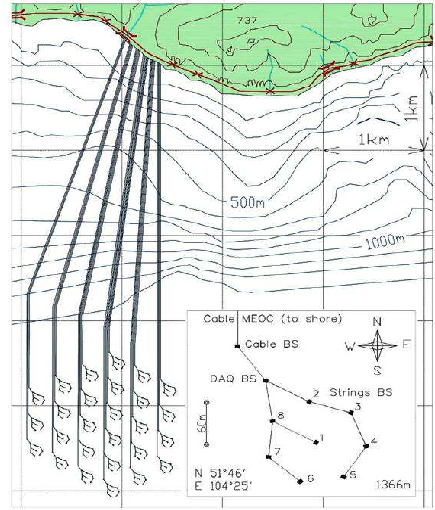}
 \end{center}
\caption{\label{fig:mapGVD} 
Figure 1: Layout of the GVD. In inner box the one cluster is shown}  
\end{figure}

\begin{figure}
\begin{center}
\includegraphics[width=0.33\textwidth,angle=-90]{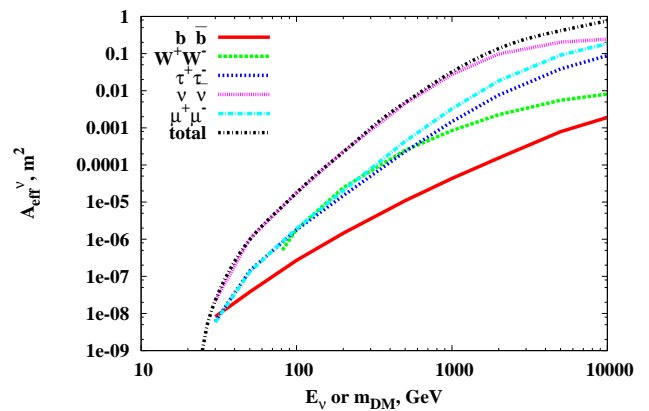}
\end{center}
\caption{\label{fig:1} 
Figure 2: Neutrino effective area of single Baikal-GVD cluster (black)
and averaged over neutrino spectra effective areas for different
annihilation channels (color).}    
\end{figure}

Stages of the GVD telescope design, its commissioning and deployment
of the first prototypes are successfully completed. Presently
there are five strings of the first demonstration cluster operating
since April 2014~\cite{GVD:2014,Avrorin:2013uyc}. The first cluster with
eight strings will be fully deployed next April 2015. The Baikal-GVD
is designed as 3D arrays of photomultiplier tubes (PMTs) each enclosed
in an optical module.
Each optical module consists of a pressure-resistant glass sphere with
43.2 cm diameter which holds OM electronics and PMT surrounded by a high
permittivity alloy cage for shielding it against the Earths
magnetic field. Large photomultiplier tube Hamamatsu R7081-100
is selected as light sensor of OM. The tube gain adjusted to about
$10^7$ and factor 10 by the first  channel of the preamplifier results
in a spectrometric channel linearity range up to about 100
photoelectrons (see~\cite{GVD:2014} for details).
The OMs are arranged on vertical load-carrying cables to form strings. 
Clusters of strings form functionally independent subarrays connected to 
shore by individual electro-optical cables as shown in
Fig.~\ref{fig:mapGVD}. Each cluster has a central string identical to
seven others distant at radius of 60 meters. The OMs are spaced
by 15 m along each string and are faced downward. They are
combined in sections on each string. In current design there are
two section on cluster. The distance between the central strings of
neighboring clusters is 300 meters. Details of data acquisition, basic
controls, methods of calibrations, hard- and soft-ware triggers can be
found in~\cite{Baikal:2011,GVD:2013,GVD:2014}. For the present study
we apply a muon trigger formed by requirements to select events with
at least 6 fired OMs on at least 3 strings within 500 ns. Here we make
a conservative estimate and suppose the worse situation with angular
resolution $4.5^\circ$ for track events. The neutrino effective area
at trigger level selection (3/6) as a function of energy for one
cluster for neutrinos from GC is presented in Fig.~\ref{fig:1} by
black line. 

Expected neutrino flux from DM annihilations in the Galaxy has the
following form
\begin{equation}
\label{eq:1}
\frac{d\phi_{\nu}}{dE} = \frac{\langle \sigma_A v\rangle}{2}\;
J_{2}(\psi)\;\frac{R_0\rho_{local}^{2}}{4\pi m_{DM}^2}\;\frac{dN_{\nu}}{dE}.
\end{equation}
Here $\langle \sigma_A v\rangle$ is annihilation cross section
averaged over DM velocity distribution, $dN_{\nu}/dE$ is neutrino (and
antineutrino) spectrum per act of annihilation. The dimensionless
quantity $J_{2}(\psi)$ is the square of the DM density in MW,
$\rho^2(r)$, integrated along the line of sight and rescaled by the
distance from GC to the Solar system $R_0$ and by the local DM density
$\rho_{local}$ as follows 
\begin{equation}
\label{eq:2}
J_2(\psi) = \int_{0}^{l_{max}}\frac{dl}{R_0}\;
\frac{\rho^2\left(\sqrt{R_0^2-2rR_0\cos{\psi}+r^2}\right)}{\rho_{local}^2},
\end{equation}
where $\psi$ is the angular distance from GC to the direction of
observation and the integration in~\eqref{eq:2} goes to $l_{max}$
which is much larger than the size of the Galaxy. There are 
several models for the DM density profile in the galaxies and, 
in particular, in the Milky Way\footnote{We do not take into account
  DM substructures which in general could modify the
  results~\cite{Berezinsky:2014wya}. }. 
Here we consider Navarro-Frenk-White 
(NFW)~\cite{Navarro:1995iw,Navarro:1996gj}, the Kravtsov
et.al.~\cite{Kravtsov:1997dp}, the Moore et.al.~\cite{Moore:1999nt} 
and Burkert~\cite{Burkert:1995yz} profiles. The last model is
currently favored by the observational data~\cite{Nesti:2013uwa}.
The profiles can be parametrized by 
\begin{equation}
\label{eq:3}
\rho(r) = \frac{\rho_0}{\left(\delta + \frac{r}{r_s}\right)^{\gamma}\left[1 +
    \left(\frac{r}{r_s}\right)^{\alpha}\right]^{(\beta-\gamma)/\alpha}},
\end{equation}
where numerical quantities are presented in Table~\ref{tab:1}.
\begin{table}
\begin{center}
\begin{tabular}{|c|c|c|c|c|c|c|}
\hline
Model & $\alpha$ & $\beta$ & $\gamma$ & $\delta$ & $r_s$, kpc &
$\rho_0$, GeV/cm$^3$ \\
\hline
NFW  & 1 & 3 & 1 & 0 & 20 & 0.3 \\
\hline
Kravtsov & 2 & 3 & 0.4 & 0 & 10 & 0.37 \\
\hline
Moore & 1.5 & 3 & 1.5 & 0 & 28 & 0.27 \\
\hline
Burkert & 2 & 3 & 1 & 1 & 9.26 & 1.88\\
\hline
\end{tabular}
\caption{\label{tab:1} Table~1: Parameters of DM density
    profiles.}  
\end{center}
\end{table}

In the case of dark matter decay in the Galaxy expected neutrino
flux is
\begin{equation}
 \label{eq:4}
\frac{d\phi_{\nu}}{dE} = \frac{1}{\tau_{DM}}\;
J_{1}(\psi)\;
\frac{R_0\rho_{local}}{4\pi m_{DM}}\;
\frac{dN_{\nu}}{dE},
\end{equation}
where $\tau_{DM}$ is DM particle lifetime and $J_1(\psi)$ is the
following integral 
\begin{equation}
\label{eq:5}
J_{1}(\psi) = \int_{0}^{l_{max}}\frac{dl}{R_0}\;
\frac{\rho\left(\sqrt{R_0^2-2rR_0\cos{\psi}+r^2}\right)}{\rho_{local}}.
\end{equation}

\begin{figure}
\begin{center}
\includegraphics[width=0.35\textwidth,angle=-90]{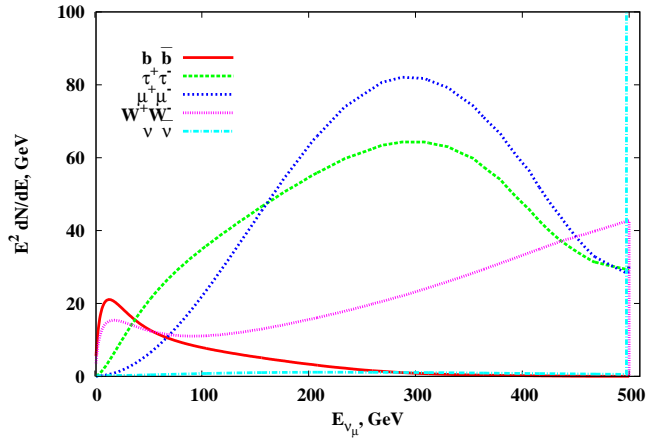}
\end{center}
\caption{\label{fig:2} 
Figure 3: Neutrino $\nu_{\mu}$ energy spectra at the Earth for
$m_{DM}=500$~GeV.} 
\end{figure}

Neutrino spectra from dark matter annihilation/decay have been taken
from~\cite{Baratella:2013fya}. We consider $b\bar{b}$,
$\tau^{+}\tau^{-}$, $\mu^{+}\mu^{-}$, $W^{+}W^{-}$ and $\nu\bar{\nu}$
channels, where in the latter case we assume flavor symmetric
annihilation. Note that 
the authors of~\cite{Baratella:2013fya} artificially modified the 
neutrino spectra for $\nu\bar{\nu}$ to be able to solve their
evolution equations which resulted to a large smearing of the
monochromatic line. We change these spectra back to their physical
width conserving the absolute norm\footnote{We are grateful to
  Marco Cirelli for the correspondence on this issue.}.
For calculation of muon-neutrino energy spectra at the Earth we use
probabilities for long-baseline oscillations. 
As neutrino oscillation parameters we use~\cite{Forero:2014bxa} the
following values: $\Delta m_{21}^2 = 7.6\cdot 10^{-5}~{\rm eV}^2$,
$\Delta m_{31}^{2} = 2.48\cdot 10^{-3}~{\rm eV}^2$, $\delta_{CP} = 0$,
$\sin^2{\theta_{12}} = 0.323$, $\sin^2{\theta_{23}} = 0.567$,
$\sin^2{\theta_{13}} = 0.0234$.
Neutrino spectra at the Earth level are presented in
Fig.~\ref{fig:2} for $m_{DM}=500$~GeV.  
We simulate neutrino propagation through the Earth and obtain muon
flux at the level of the detector as it was described
in~\cite{Boliev:2013ai,Avrorin:2014swy}.

To estimate the sensitivity of GVD to neutrinos from dark matter
annihilations in GC we choose a search region as a cone around the
direction towards the GC with half angle $\psi_0$. The expected number
of signal events in the search region for the livetime $T$ is
estimated as 
\begin{equation}
\label{eq:6}
N(\psi_0) = T\frac{\langle\sigma_A v\rangle R_0\rho_{local}^{2}}{8\pi
  m_{DM}^2}J_{2, \Delta\Omega}\int
dE S(E)\frac{dN_{\nu}}{dE}.
\end{equation}
Here $S(E)$ is neutrino effective area of the telescope. In
Fig.~\ref{fig:1} along with effective area of one cluster (black) we
show effective areas averaged over given neutrino spectrum
$\frac{dN_{\nu}}{dE}$ for neutrinos coming from GC.  
The factor $J_{2, \Delta\Omega}$ is obtained by averaging of
$J_2(\psi)$ over the search region 
\begin{equation}
J_{2, \Delta\Omega} = \int d(cos{\psi})d\phi J_{2}(\psi)\epsilon(\psi,\phi),
\end{equation}
where $\epsilon(\psi, \phi)$ is visibility of the corresponding point
on the sky by the telescope.

\begin{figure}
\begin{center}
\includegraphics[width=0.45\textwidth]{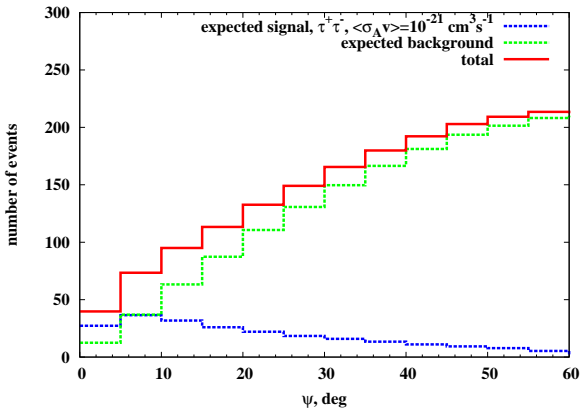}
\caption{\label{fig:3} 
Figure 4: Distribution of background and signal events in angular
distance from GC.}  
\end{center}
\end{figure}

The atmospheric neutrino background is estimated from MC simulations
with trigger conditions described above. The average total number of
background events coming from low hemisphere for one year is
expected to be 4300. Distribution of background events in angular
distance from GC is shown in Fig.~\ref{fig:3} in green. 
Angular distribution of signal events for $\tau^+\tau^-$ annihilation
channel with $m_{DM}=200$~GeV and $\langle\sigma_A
v\rangle=10^{-21}$cm$^3$s$^{-1}$ is also shown in this Figure in
blue. Using the differences in their angular distributions we
optimize the size of the search region with respect to cone half-angle
in the following standard way. Assuming absence of the signal we
construct the ratio $\bar{N}^{90}(\psi_0)/\sqrt{N_B}$ where
$\bar{N}^{90}(\psi_0)$ is 90\% CL upper limit on number of events in
the given cone averaged over number of observed events with
Poisson distribution and $N_B$ is expected number of background 
events. We maximize it with respect to $\psi_0$ and obtained optimized
values of cone half angles vary from 9$^\circ$ for hard
($\nu\bar{\nu}$) channels and large DM masses to $21^\circ$ for soft
channels ($b\bar{b}$) and small values of $m_{DM}$.  

\begin{figure}
\begin{center}
\includegraphics[width=0.34\textwidth,angle=-90]{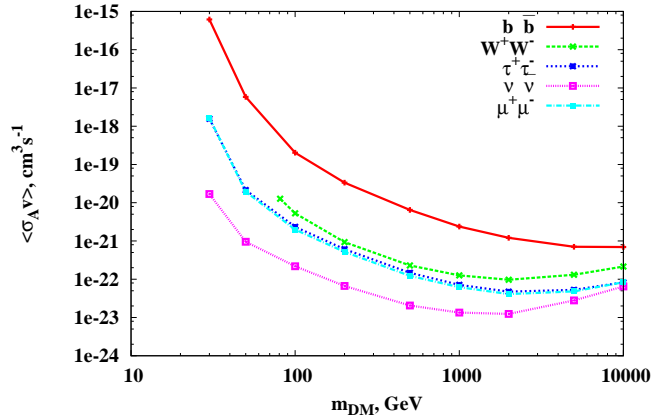}
\caption{\label{fig:4} 
Figure 5: Sensitivity of GVD to $\langle\sigma v\rangle$ for
one year for different annihilation channels.}  
\end{center}
\end{figure}

\begin{figure}
\begin{center}
\includegraphics[width=0.34\textwidth,angle=-90]{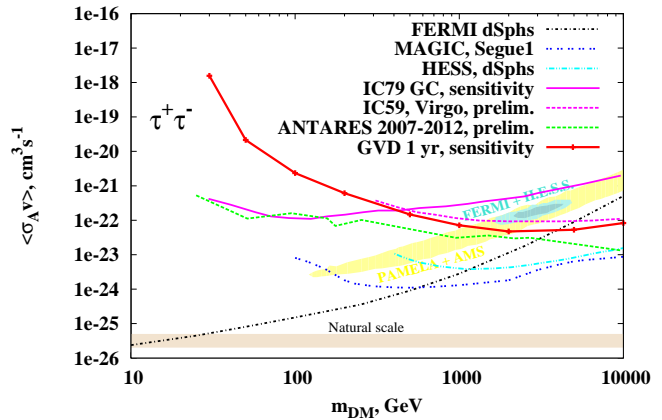}
\caption{\label{fig:4_1} 
Figure 6: Sensitivity of GVD to $\langle\sigma v\rangle$ in
comparison with other experiments.}  
\end{center}
\end{figure}

The expected upper bounds on dark matter annihilation cross section
have been obtained from~\eqref{eq:6} where $N(\psi_0)$ were replaced
by $\bar{N}^{90}(\psi_0)$. Further, we include realistic estimates for
detection efficiency $\epsilon=0.6$ and systematic uncertainties
$\eta=50\%$, into the bounds as
$\bar{N}^{90}(\psi_0)/\left(\epsilon(1-\eta)\right)$. The sensitivity
to DM annihilation cross sections for NFW density profile is shown in
Figs.~\ref{fig:4} and~\ref{fig:4_1} 
in comparison with the results/sensitivities
of other experiments FERMI~\cite{Ackermann:2013yva},
MAGIC~\cite{Aleksic:2013xea}, H.E.S.S.~\cite{Abramowski:2014tra}, 
IceCube~\cite{Aartsen:2013mla,Aartsen:2013dxa},
ANTARES~\cite{Zornoza:2014cra} and with the results of DM
interpretation of positron excess~\cite{Meade:2009iu}.
Also our estimates show possible improvement in the bounds for better
angular resolution of 1$^\circ$ up to $8-10\%$. 

\begin{figure}
\begin{center}
\includegraphics[width=0.34\textwidth,angle=-90]{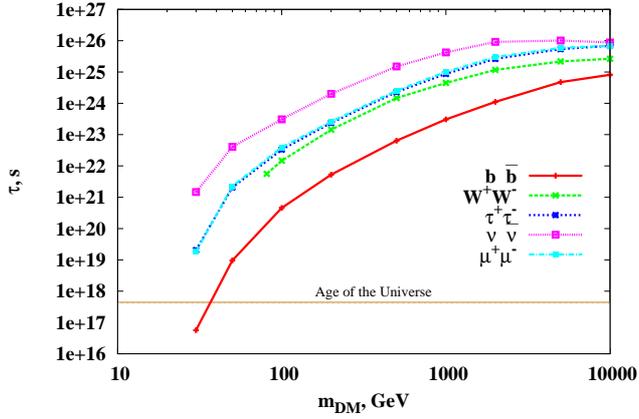}
\caption{\label{fig:5} 
Figure 7: GVD-Baikal sensitivity to $\tau_{DM}$ for $T=1$~yr.} 
\end{center}
\end{figure}

To get expected low bound on the DM lifetime $\tau_{DM}$ we use
the search regions obtained before and apply Eq.~\eqref{eq:5}. We plot 
  GVD sensitivity to DM decays for NFW profile in Fig.~\ref{fig:5}  
for chosen set of decay channels. We note that the angular
distribution of signal events in the case of DM decay is different
from the case of its annihilation. We perform new optimization with 
respect to the search region for decaying DM and the optimal values of
$\psi_0$ are obtained to be about $70^\circ$ which is well beyond the
neighbourhood of GC. From this analysis we expect an improvement by
factor $2-3$ in the bounds on $\tau_{DM}$ with searches in the whole
Galactic halo.  

\begin{figure}
\begin{center}
\includegraphics[width=0.34\textwidth,angle=-90]{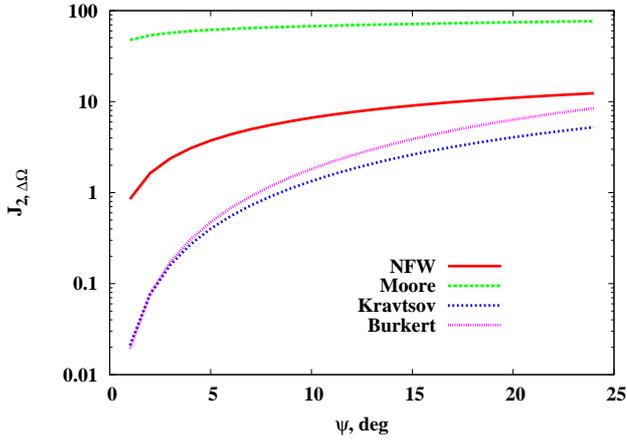}
\caption{\label{fig:6} 
Figure 8: $J_{2}$-factors for different models of dark matter density
profile.} 
\end{center}
\end{figure}

There are several theoretical uncertainties in the number of signal
events related to neutrino oscillation parameters, neutrino-nucleon
cross section etc. However, the most important of them is the 
uncertainty related to our lack of knowledge of DM density
profile near GC. To illustrate this issue in Fig.~\ref{fig:6} 
we plot $J_{2}$-factors for different models of DM distribution. We 
see that the difference can reach several orders of magnitude.
Note that the factor $J_{1}$ is less model dependent.

To summarize, we studied the GVD sensitivity to DM
annihilations/decays in GC for 1 year of livetime at trigger selection
level. The expected bounds for realistic efficiency and
systematic uncertainties reach values about $10^{-23}$~cm$^3$s$^{-1}$
for dark matter annihilation cross section and $10^{26}$~s for DM
lifetime in the most energetic $\nu\bar{\nu}$ channel.   

The work of S.V.~Demidov and O.V.~Suvorova was supported by the RSCF
grant 14-12-01430.

\end{document}